\documentclass[hyper]{JHEP3}
\usepackage{graphicx}
\usepackage{cite}
\def\beq{\begin{equation}}
\def\eeq{\end{equation}}
\def\beqn{\begin{eqnarray}}
\def\eeqn{\end{eqnarray}}
\def\nn{\nonumber\\}

\title{Mass determination in sequential particle decay chains\footnote{Research supported in part by the
UK Science and Technology Facilities Council.}}

\author{Bryan Webber\\
        Cavendish Laboratory, J.J.\ Thomson Avenue, Cambridge, UK\\
        E-mail: \email {webber@hep.phy.cam.ac.uk}
        }

\received{\today}               
\accepted{\today}               

\preprint{Cavendish-HEP-09/13}

\abstract{
A simple method is proposed for determining the masses of new particles in collider
events containing a pair of decay chains (not necessarily identical) of the form
$Z\to Y+1$, $Y\to X+2$, $X\to N+3$, where 1,2 and 3 are visible but $N$ is not.
Initial study of a possible supersymmetric case suggests that the method can
determine the four unknown masses in effectively identical chains with good
accuracy from samples of a few tens of events. 
}


\begin{document} 

\section{Introduction}
\label{sec:intro}
The discovery of new physics beyond the Standard Model is a primary objective of experiments at
the Large Hadron Collider and other future colliders.  In many models of BSM physics, a rich spectrum
of new particles is predicted in the mass range accessible at the LHC.  Many of these particles are
weakly interacting and have quantum numbers that favour sequential decays into chains of other
new particles plus visible jets and/or leptons.  Typically the endpoint of the chain is a stable
invisible particle that is a dark matter candidate.  Classic examples are the squark decay chain in
supersymmetric models,
\beq\label{eq:chain_susy}
\tilde q\to \tilde\chi^0_2 + q\,,\;\tilde\chi^0_2 \to \tilde\ell^\pm+\ell^\mp\,,\;
\tilde\ell^\pm\to\tilde\chi^0_1+\ell^\pm\;,
\eeq
where the neutralino $\tilde\chi^0_1$ is the lightest supersymmetric particle (LSP),
and the excited quark decay in models with universal extra dimensions,
\beq\label{eq:chain_ued}
q^*\to Z^* + q\,,\; Z^* \to \ell^{*\pm}+\ell^\mp\,,\;
\ell^{*\pm}\to\gamma^*+\ell^\pm\;,
\eeq
where the photon excitation $\gamma^*$ is the lightest Kaluza-Klein particle (LKP).

Determining the masses of the new particles in such decay chains, especially the
dark matter candidate, is clearly of great importance.  Many approaches to this problem have been
proposed~\cite{Hinchliffe:1996iu,Paige:1997xb,Lester:1999tx,Bachacou:1999zb,Hinchliffe:1999zc,
Tovey:2000wk,Allanach:2000kt,Barr:2003rg,Nojiri:2003tu,Kawagoe:2004rz,Gjelsten:2004ki,
Gjelsten:2005aw,Miller:2005zp,Lester:2006yw,Gjelsten:2006as,Gjelsten:2006tg,Cheng:2007xv,
Lester:2007fq,Cho:2007qv,Gripaios:2007is,Barr:2007hy,Cho:2007dh,Ross:2007rm,Nojiri:2007pq,
Huang:2008ae,Nojiri:2008hy,Tovey:2008ui,Cheng:2008mg,Serna:2008zk,
Bisset:2008hm,Barr:2008ba,Kersting:2008qn,Nojiri:2008vq,Alwall:2008va,Cho:2008tj,Cheng:2008hk,
Burns:2008va,Barr:2008hv,Konar:2008ei,Kersting:2009ne,Costanzo:2009mq,Papaefstathiou:2009hp,
Burns:2009zi,Cheng:2009fw,Serna:2009gw,Matchev:2009iw,Han:2009ss}, based mainly on the
measurement of endpoints or other features in the distributions of invariant masses or specially
constructed observables, or on explicit solution for the unknown masses using multiple events.

The present paper investigates a somewhat different approach which is particularly suited to
processes in which there are two three-step decay chains of the form (\ref{eq:chain_susy}) or  (\ref{eq:chain_ued}).  In principle the chains need not be the same, although in practice it would
be too much to expect to determine the eight masses involved in non-identical chains.  We shall
see that for practically identical chains, such as first- and second-generation squark pair production
and decay as in (\ref{eq:chain_susy}), determination of the four sparticle masses and reconstruction
of the LSP momenta appears possible with reasonable numbers of events.

\section{Method}
\begin{figure}\begin{center}
\includegraphics[width=70mm]{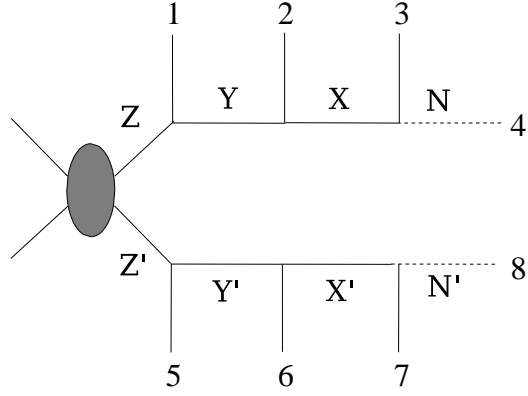}
\caption{Event topology}\label{fig:event}
\end{center}\end{figure}
Consider the double decay chain in fig.~\ref{fig:event}.
The 4-momenta in the upper chain should satisfy
\beqn\label{eq:upper}
(p_1+p_2+p_3+p_4)^2 &=& M_Z^2\nn
(p_2+p_3+p_4)^2 &=& M_Y^2\nn
(p_3+p_4)^2 &=& M_X^2\nn
p_4^2 &=& M_N^2
\eeqn
Leaving aside the last equation, the others give three linear constraints
on the invisible 4-momentum $p_4$:
\beqn\label{eq:upconstr}
-2p_1\cdot p_4 &=& M_Y^2-M_Z^2+2p_1\cdot p_2 +2p_1\cdot p_3+m_1^2\equiv S_1\nn
-2p_2\cdot p_4 &=& M_X^2-M_Y^2+2p_2\cdot p_3 +m_2^2\equiv S_2\nn
-2p_3\cdot p_4 &=& M_N^2-M_X^2+m_3^2\equiv S_3
\eeqn
Similarly for the lower chain
\beqn\label{eq:loconstr}
-2p_5\cdot p_8 &=& M_{Y'}^2-M_{Z'}^2+2p_5\cdot p_6 +2p_5\cdot p_7+m_5^2\equiv S_5\nn
-2p_6\cdot p_8 &=& M_{X'}^2-M_{Y'}^2+2p_6\cdot p_7 +m_6^2\equiv S_6\nn
-2p_7\cdot p_8 &=& M_{N'}^2-M_{X'}^2+m_7^2\equiv S_7
\eeqn
We also have the missing transverse momentum constraints
\beqn\label{eq:trans}
p_4^x+p_8^x &=& p^x_{\rm miss}\equiv S_4\nn
p_4^y+p_8^y &=& p^y_{\rm miss}\equiv S_8
\eeqn
Let us make an 8-vector of the invisible 4-momenta,
\beq
{\bf P} = (p_4^x,p_4^y,p_4^z,E_4,p_8^x,p_8^y,p_8^z,E_8)
\eeq
Then we have
\beq
{\bf A P} ={\bf S}
\eeq
where ${\bf A}$ is the 8$\times$8 matrix
\beq\label{eq:Adef}
{\bf A} = 2\left(\begin{array}{cccccccc}
p_1^x & p_1^y & p_1^z & -E_1 & 0 & 0 & 0 & 0 \\
p_2^x & p_2^y & p_2^z & -E_2 & 0 & 0 & 0 & 0 \\
p_3^x & p_3^y & p_3^z & -E_3 & 0 & 0 & 0 & 0 \\
 1/2 & 0 & 0 & 0 & 1/2 & 0 & 0 & 0\\
 0 & 0 & 0 & 0 & p_5^x & p_5^y & p_5^z & -E_5\\
 0 & 0 & 0 & 0 & p_6^x & p_6^y & p_6^z & -E_6\\
 0 & 0 & 0 & 0 & p_7^x & p_7^y & p_7^z & -E_7\\
 0 & 1/2 & 0 & 0 & 0 & 1/2 & 0 & 0
\end{array}\right)
\eeq
Furthermore ${\bf S}$ may be written as
\beq
{\bf S} = {\bf B M} + {\bf C}
\eeq
where ${\bf M}$ is the vector of masses-squared to be determined,
\beq\label{eq:Mdef}
{\bf M} = (M_Z^2,M_Y^2,M_X^2,M_N^2,M_{Z'}^2,M_{Y'}^2,M_{X'}^2,M_{N'}^2)\;,
\eeq
\beq
{\bf B} = \left(\begin{array}{cccccccc}
-1 & 1 & 0 & 0 & 0 & 0 & 0 & 0\\
0 & -1 & 1 & 0 & 0 & 0 & 0 & 0\\
0 & 0 & -1 & 1 & 0 & 0 & 0 & 0\\
0 & 0 & 0 & 0 & 0 & 0 & 0 & 0 \\
0 & 0 & 0 & 0 &-1 & 1 & 0 & 0\\
0 & 0 & 0 & 0 & 0 & -1 & 1 & 0\\
0 & 0 & 0 & 0 & 0 & 0 & -1 & 1\\
0 & 0 & 0 & 0 & 0 & 0 & 0 & 0
\end{array}\right)
\eeq
and
\beqn\label{eq:Cdef}
{\bf C} &=&
(2p_1\cdot p_2 +2p_1\cdot p_3+m_1^2, 2p_2\cdot p_3 +m_2^2, m_3^2,p^x_{\rm miss},\nn
&&\;\,2p_5\cdot p_6 +2p_5\cdot p_7+m_5^2,2p_6\cdot p_7 +m_6^2,m_7^2,p^y_{\rm miss})\;.
\eeqn
Hence the solution for the invisible 4-momenta is
\beq
{\bf P} ={\bf A}^{-1}{\bf S} = {\bf D M} + {\bf E}
\eeq
where ${\bf D} = {\bf A}^{-1}{\bf B}$ and  ${\bf E} ={\bf
  A}^{-1}{\bf C}$.

Since ${\bf A}_n$  for each event $n$ is a sparse matrix it is
easy to invert it and hence to obtain and store ${\bf D}_n$ and ${\bf
  E}_n$ (72 real numbers) for each event.  Then for every hypothesis
for the unknown masses ${\bf M}$ we immediately get a unique
solution for the invisible 4-momenta in each event, ${\bf P}_n$.
For the correct mass hypothesis, these satisfy the quadratic constraints
\beqn
(p_4^2)_n &=& (P_4^2-P_1^2-P_2^2-P_3^2)_n =M_N^2\nn
(p_8^2)_n &=& (P_8^2-P_5^2-P_6^2-P_7^2)_n =M_{N'}^2
\eeqn
We can therefore measure the goodness of fit for the mass hypothesis ${\bf M}$ by
the quantity\footnote{The symbol $\xi^2$ rather than $\chi^2$ reminds us that it has
no probabilistic interpretation.  However, in an experimental analysis event-to-event
variations in momentum uncertainties could be propagated from eqs.~(\ref{eq:Adef})
and (\ref{eq:Cdef}) into (\ref{eq:xidef}) to give greater weight to events with higher
precision, thus defining a quantity more like $\chi^2$.  My thanks to Ben Gripaios
for this suggestion.}
\beq\label{eq:xidef}
\xi^2({\bf M}) = \sum_n \left[ (p_4^2)_n - M_N^2\right]^2 + \sum_n \left[ (p_8^2)_n
    - M_{N'}^2\right]^2
\eeq
The method is then to find the best-fit hypothesis for the masses by minimizing $\xi^2$.
Note that this quantity tests the goodness of fit to all the masses equally, since for
example it follows from eq.~(\ref{eq:upconstr}) that
\beq
(p_3+p_4)^2_n - M_X^2 = (p_4^2)_n - M_N^2
\eeq
and similarly for all the other unknown masses.

To account for combinatorial ambiguities, we must evaluate $\xi^2$ for all permitted
particle combinations for each event.   Notice however that a different combination
within one chain corresponds to a permutation of the rows of the matrix ${\bf A}$. 
Therefore the inverse is given by the same permutation of the columns of ${\bf A}^{-1}$,
and no extra inversions or matrix storage are necessary.  In the case that the mass
difference between particles 2 and 3 is negligible (as for dileptons), the vector ${\bf C}$
is invariant under their exchange; similarly for 6 and 7.  Thus only combinations that
exchange particles between chains require extra data storage.

For every mass hypothesis, we should use the lowest value of $\xi^2$ amongst all
the allowed combinations for each event. At the best-fit point, this also shows
which combination is most likely to be the correct one.  Then the corresponding
reconstructed momenta can be used, for example, to test spin hypotheses for
the particles involved.

This method is closest in spirit to that of refs.~\cite{Cheng:2008mg,Cheng:2009fw}, in which
pairs of events with the same decay chains are used to solve explicitly for the unknown masses.
As this is a non-linear problem, each pair yields multiple solutions, which are narrowed down to
the correct one as more pairs are solved.  The advantage of the present method is that each
event contributes independently and additively to the goodness-of-fit function (\ref{eq:xidef}),
which is obtained by a simple linear computation for any mass hypothesis.  The problem then
reduces to the familiar one of function minimization.  The additivity property also means that
combining event samples and the statistical interpretation of results become more
straightforward.  Furthermore the results from other methods such as edge analyses can
easily be included as constraints on the mass parameters in the minimization search.

\section{Results}

As an illustration of the method, it was applied to the process of squark-pair
production at the LHC ($pp$ at 14 TeV centre-of-mass energy).  The SUSY mass
spectrum and decay branching ratios were taken to be those of CMSSM point
SPS 1a~\cite{Allanach:2002nj}.  At this point the SUSY production cross section
at the LHC is about 50 pb and there is a good probability for squark production
and decay into quark jets and dileptons via (\ref{eq:chain_susy}).
Some of the squarks are produced directly and some come from
gluino decay; the production mechanism affects their momentum
and rapidity distributions but is otherwise irrelevant for our purposes.
Decays of the two squarks into unlike dileptons  ($e^+e^-\mu^+\mu^-$)
were selected to limit the number of allowed combinations of jets and near
and far leptons to eight, necessitating the storage of 144 real numbers for
each event as explained above.

Third-generation squarks were excluded, as their different masses prevent a good
fit with a single squark mass.  Experimentally, this would involve vetoing events
with a tagged $b$-jet.  At SUSY point SPS 1a only left-squarks have significant
branching ratios into the mode (\ref{eq:chain_susy}) and so the
left-right squark mass splitting is not a problem here.  Therefore a
four-parameter fit with $M_{i+4}=M_i$ in eq.~(\ref{eq:Mdef}) is
appropriate.

\begin{figure}\begin{center}
\includegraphics[width=100mm]{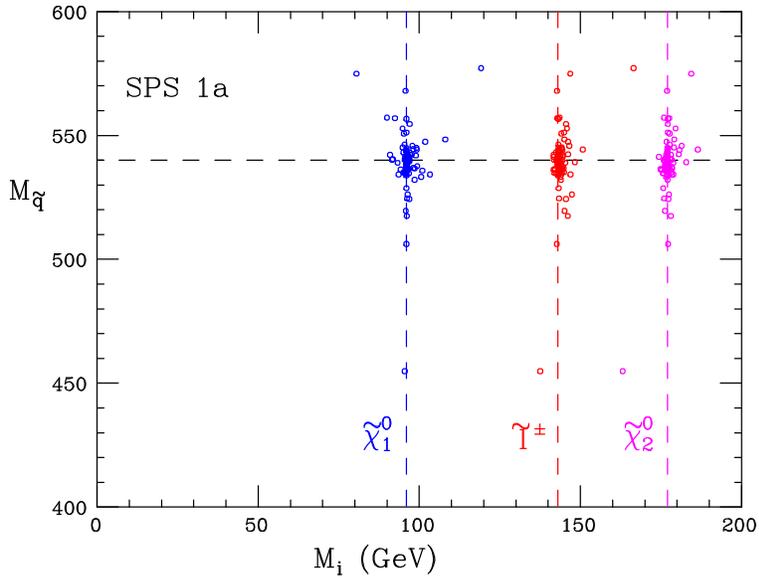}
\caption{Best-fit results for SUSY point SPS 1a.  Each point represents a
sample of 25 events.  The dashed lines show the true masses.}\label{fig:SPS1a}
\end{center}\end{figure}

Figure~\ref{fig:SPS1a} shows the best-fit results for 100 Monte Carlo samples
of 25 events each, selected as described above.  A sample of 25 such events
corresponds to an integrated luminosity of about 3 fb$^{-1}$.
The events were generated with {\small HERWIG} version
6.510~\cite{Corcella:2000bw,Corcella:2002jc,Moretti:2002eu} and the jet and
lepton momenta used in the analysis were at parton level (after parton
showering but before hadronization), with perfect jet reconstruction
and no momentum smearing in this figure. The missing transverse
momentum was taken to be that of the LSPs alone, again without smearing.
However,  {\small HERWIG} smears all unstable particle masses with the
appropriate Breit-Wigner distributions, an effect that is significant
for the squarks ($\Gamma_{\tilde q}\sim 5$ GeV) but negligible for
the sleptons and neutralinos.

We see that in this idealized situation the four new particle masses are
usually quite well determined.  As summarized in the first row of
table~\ref{tab:fits}, the r.m.s.\ variation in the estimated mass is 20 GeV
for the squark and around 10 GeV for the slepton and neutralinos,
with mean values within 1 or 2 GeV of the true ones.
The best-fit combination is the correct one (at the best-fit point)
in 72\% of events. The fits are not perfect, and there are incorrect
choices of combination, because of the intrinsic differences amongst the
squark and slepton masses and the Breit-Wigner smearing.

\begin{table}
  \begin{center}    
    \begin{tabular}{|r|c|c|c||l|l|l|l|}
      \hline
 $\delta p/p$ & $\xi^2_{\rm max}$ &  $f_\xi$ & $f_{\rm cor}$ &
$M_{\tilde q}$ (540) & $M_{\tilde\chi^0_2}$ (177) & $M_{\tilde\ell}$ (143) & $M_{\tilde\chi^0_1}$ (96)\\
\hline\hline
 0 & $\infty$  & 100\% & 72\%    & $538\pm 20$ & $176\pm 12$ & $143\pm 7$ & $95\pm 10$\\
 0 & 100   &   80\% & 76\%          & $539\pm 7$ & $177\pm 1$   & $144\pm 1$   & $96\pm 2$\\
 5\% & $\infty$ & 100\% & 52\% & $534\pm 28$ & $176\pm 11$ & $143\pm 10$ & $95\pm 13$\\
 5\% & 100  &   57\% & 55\%       & $539\pm 9$ & $178\pm 3$   & $144\pm 2$   & $96\pm 4$\\
 10\% & $\infty$ & 100\%&40\% & $522\pm 37$ & $171\pm 18$ & $140\pm 17$ & $88\pm 26$\\
 10\% &  200  &   42\% & 43\%    & $530\pm 22$ & $173\pm 12$  & $140\pm 12$ & $89\pm 20$\\
     \hline
    \end{tabular}
  \end{center}
  \caption{
\label{tab:fits}
Fitted masses and r.m.s.\ variations for samples of 25 events.  The true average masses
are shown in the heading (all in GeV).  The quantity $f_\xi$ is the fraction of samples
surviving the $\xi^2$ cut, while $f_{\rm cor}$ is the fraction of events with the correct
best-fit combination.
}
\end{table}

The search for the best fit is rather tricky because the $\xi^2$-surface is
not smooth, owing to sudden changes in the best-fit combinations as the
mass parameters are varied.  For this study, the {\small SIMPLEX} method
in {\small MINUIT}~\cite{{James:1975dr}} was used.  The surface would be
smooth if one added the  $\xi^2$ contributions of all combinations, but
then the sensitivity to the correct solution is reduced and biases
are introduced by the huge contributions of wrong combinations.

The mass resolution can be improved by eliminating data sets with
large best-fit values of $\xi^2$.  For example, if we require our 25
event sample to have total $\xi^2<100$ in units of (100 GeV)$^4$, then
80\%  of the samples survive and the fluctuations in the
fitted masses are reduced as indicated in the second row of table~\ref{tab:fits}.

\begin{figure}\begin{center}
\includegraphics[width=100mm]{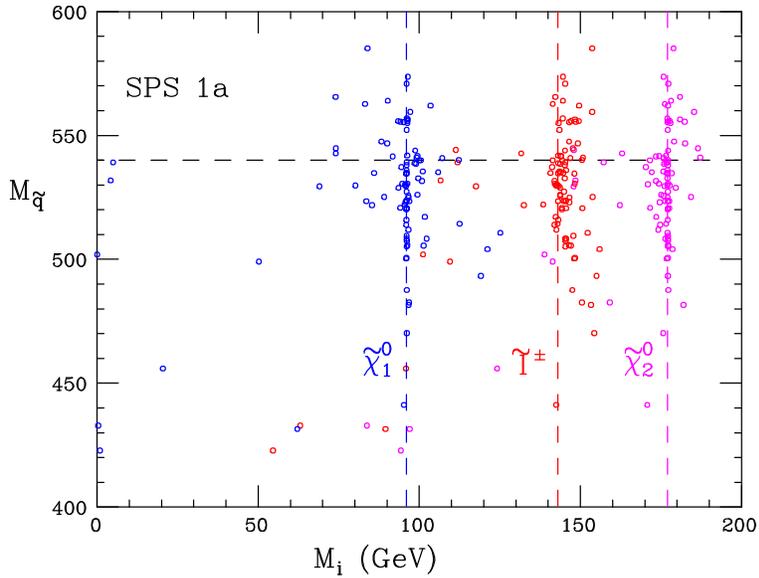}
\caption{As in figure 2, but with momentum smearing $\delta p/p = 10\%$.}\label{fig:SPS1a_smear10}
\end{center}\end{figure}

As a rough indication of the possible effects of hadronization, reconstruction errors and
detector resolution, the jet and lepton momenta and also the missing momenta were smeared
with a gaussian distribution of r.m.s.\ width $\delta p/p = 5\%$ or 10\%, with the results
shown in the lower rows of table~\ref{tab:fits} and, for 10\% smearing, in
fig.~\ref{fig:SPS1a_smear10}.

The fluctuations in the fitted masses naturally increase with increasing smearing,
and the fraction of correct best-fit combinations decreases.  There is a slight bias of the
mass estimates towards lower values due to momentum smearing.  As in the unsmeared
case, a cut on $\xi^2$ reduces the fluctuations at the expense of rejecting a fraction of the
event samples.  This means, for example, that in a single experiment with 25 events and 10\%
smearing there is about a 58\% chance that $\xi^2 > 200$, in which case the estimates of the
masses are likely to be poorer than those shown in the last row of the table.

Since the individual events are statistically independent, the fluctuations in the best-fit
mass estimates decrease inversely as the square root of the size of the event sample.
Any systematic  biases in these estimates would not decrease with improving statistics,
but the results with momentum smearing suggest that such effects should be small.
They could be corrected using more detailed Monte Carlo simulations if present in a
real experimental analysis.

One possible source of bias would be the neglect of jet invariant mass in the reconstruction
of the quark jets.  In the present study, the effect of this was investigated by rescaling either the
jet momentum or the jet energy to give zero jet mass.  In the former case there was little effect,
but energy rescaling resulted in a downward bias of about 25 GeV in the estimated squark mass
and around 5 GeV in the other masses.   

\section{Conclusions}

The method of mass determination presented above is simple to apply and looks
promising for the class of processes studied here.  It can readily be extended to more
complicated final states involving different or longer decay chains.  Combinatorial
background does not appear to be a serious problem although other backgrounds and
the effects of additional jets due to QCD radiation remain to be investigated.  These would
be best studied in the framework of full simulations including detector effects.  Comparative
studies along these lines for this and other mass determination methods are in
progress\cite{leshouches}.

A variant of this method can also be applied to events with one three-step decay chain
like (\ref{eq:chain_susy}) or (\ref{eq:chain_ued}) and one shorter (two-step) chain, for example
\beq\label{eq:2chain_susy}
\tilde g\to \tilde q'+ \bar q'\,,\;\tilde q' \to \tilde\chi^0_1+ q'\;,
\eeq
or analogously
\beq\label{eq:2chain_ued}
g^*\to q'^* + \bar q'\,,\; q'^* \to\gamma^*+ q'\;.
\eeq
In this case the four constraints (\ref{eq:upper}) on the longer chain can be solved for
the invisible 4-momentum $p_4$, with a two-fold ambiguity since one constraint is now
quadratic.  The kinematics of the shorter chain can then be reconstructed
and the goodness of fit (\ref{eq:xidef}) computed for both solutions.  Choosing the
solution with the better fit for each mass hypothesis, one can now proceed as in the
case of two three-step chains.  Further discussion of this case will be presented
in a later paper\cite{inprep}.

\section*{Aknowledgements}
I am grateful for helpful discussions with Ben Allanach, Ben Gripaios, Chris Lester, Bob McElrath
and Are Raklev, and for the hospitality of the CERN Theory Group during part of this work.


\end{document}